\documentstyle[pra,aps,preprint]{revtex}
\begin{document}
\draft
\title{Quantum Transport in Two-Channel Fractional Quantum Hall Edges}
\author{K. Imura and N. Nagaosa   }
\address{Department of Applied Physics, University of Tokyo,
Bunkyo-ku, Tokyo 113, Japan}
\date{\today}
\maketitle

\begin{abstract}
We study the effect of backward scatterings in the tunneling
at a point contact between the edges of a second level
hierarchical fractional quantum Hall states.
A universal scaling dimension of the tunneling
conductance is obtained only when both of the edge channels
propagate in the same direction.
It is shown that the quasiparticle tunneling picture
and the electron tunneling picture give different scaling
behaviors of the conductances, which indicates the existence
of a crossover between the two pictures.
When the direction of two edge-channels are opposite, e.g.
in the case of MacDonald's edge construction for the $\nu=2/3$
state, the phase diagram is divided into two domains giving
different temperature dependence of the conductance.

\end{abstract}


\narrowtext

\section{Introduction}
\noindent
It has been discovered that
the edge mode of a fractional quantum Hall (FQH) system
is described as a chiral Tomonaga-Luttinger (TL) liquid ~\cite{Wen90}.
When one makes a point contact between the edges,
backward scatterings become possible only near that point
contact. Such a system is expected to be described as a TL
model with a scattering potential at $x=0$ ~\cite{Kane92,Furusaki93}.
For a $\nu=1/3$ state, the model predicts a $T^4$ dependence
of the tunneling conductance near the zero temperature
~\cite{Wen91b,Wen92,Fendley95} which is consistent with
the recent experiment~\cite{Milliken96}.
The result does not rely upon whether we start the renormalization
group (RG) analysis with the picture where quasiparticles
tunnel between the edges through the bulk FQH liquid
(quasiparticle tunneling picture) or
the picture where electrons tunnel through the vacuum
between the edges of two condensates separated by the
tunnel effect (electron tunneling picture)~\cite{Moon93}.

To describe the bulk FQH liquids with a filling factor
$\nu \neq {1\over 2k+1}$ hierarchy constructions are introduced
~\cite{Haldane83,Halperin84,Blok90a,Jain89a,Jain89b,Blok90b}.
At a given filling one can easily propose many
hierarchical constructions.
When we apply the above discussion on the edge mode to the
hierarchy states, we encounter a completely novel situation.
Let us consider a two-terminal geometry where the bulk FQH
liquids between the left and right terminals have upper and lower
edges.
Each of the edge branches consists of more than two channels.
In this case the tunneling at apoint contact as well as the
interaction between the channels become important for the
quantum transport at the edges.
For $\nu=2/5$ and 2/3
there are two channels of the edge modes ~\cite{Wen92}
which are specified by a set of filling factors
($\nu_1,\nu_2$).
However the possibility
that there are more edge channels cannot be excluded
~\cite{Chklovski95,Beenakker90}.

For $\nu=2/3$ two constructions with $(\nu_1,\nu_2)=(1,-1/3)$ and
(1/3,1/3) have been proposed, which correspond respectively to
sharp~\cite{MacDonald90} and smooth confinement~\cite{Beenakker90}.
In the former (latter) construction two edge modes in a same
branch propagate in the opposite (same) direction.
Recent numerical calculations based on the composite Fermion
approach in the Hartree approximation confirmed that there
is a transition from the MacDonald's (former) picture to the
Beenakker's (latter) one as the confinement is relaxed
~\cite{Chklovski95,Brey94}.

So far we have not taken into account the spin degrees of freedom since
we have assumed that all the spins are polarized.
However finite-size calculations of the Coulomb energy show
that the unpolarized spin configuration is preferred
when $\nu=2/5$ or $2/3$ ~\cite{Maksym89}.
The result is consistent with the experiment in tilted
magnetic fields ~\cite{Syphers88}.
Nevertheless we can realize the spin-polarized FQH states
at arbitrary filling by applying the strong parallel magnetic
field.

Wen has shown that the FQH state has gapless boundary
excitations using the gauge invariance of the bulk
effective theory ~\cite{Wen92}.
In his derivation, low-energy effective action 
for the edge mode is determined only from the
gauge invariance.
In the appendix we give an alternative derivation.
We describe the edge mode as a self-induced eigenmode
of the bulk FQH state ~\cite{Nagaosa94}.
It turns out that the existence of Maxwell terms
in the dual action of the bulk effective theory
is essential for our derivation.

We study the effects of backward scatterings at a point
contact using the effective action generalized for
the two-channel edge modes.
We concentrate on the analysis of $\nu=2/5$ and 2/3 FQH edges
but our formulation can be applied to any two-channel FQH
edges.
We assume that the interaction between the electrons is
short-ranged.
The interaction between the upper and lower branches are neglected.

We consider both of the limits where the scattering potential
is very weak and very strong.
In the weak potential limit we start the RG analysis
from the quasiparticle tunneling picture and study the
scaling behavior of the deviation of the conductance
from the quantized value.
The situation corresponds to a "high" temperature.
As the potential becomes strong, the tunneling through
the potential barrier becomes rare and we can employ
the dilute instanton gas approximation (DIGA).
As the potential further increases, it is better to start
with the electron tunneling picture.
The question is whether or not
the scaling behaviors of the tunneling conductance are
smoothly connected from one picture to the other. 

The plan of the paper is the followings.
In Sec. II the model for the edge mode is introduced.
In Sec. III the quasiparticle tunneling picture is
explained, and the RG analysis in the weak and strong
potential limit is given. Electron tunneling picture is
studied in Sec. IV. Sec. V is devoted to discussion
and conclusions.

\section{Model}
\noindent
Our model describes a two-terminal Hall bar geometry where
a two-dimensional electron system between the left and right terminals
has upper and lower edges. Low-energy excitations of the system are
the modes which are localized near the edges since bulk FQH state has
an energy gap. In our model, edge modes of both upper and lower branches
consist of two channels (1-channel and 2-channel). In the following,

\noindent
(a) We assume that long-range Coulomb interactions are screened away
and electrons in the FQH edges have only intra-branch short-range
interactions.

\noindent
(b) We consider only the backward scatterings at a point contact, since
forward scatterings are relevant but do not open an energy gap
~\cite{Wen91a},
and since backward scatterings except at a point contact are forbidden
as a result of momentum conservation.

\noindent
(c) We also assume that the spins are fully polarized.
This may not be true for $\nu=2/5$ and $2/3$. This issue will be discussed in
future publications.

\subsection{Effective action for the tunneling}
\noindent
The Lagrangian density (A7) in the appendix is the starting point of our
following analysis. We generalize it to the two-channel case as
\begin{equation}
{\cal L_0}=\sum_{I=1,2}
\left[
\frac{c_I}{8\pi\nu_I}
\left\{
\left(\frac{\partial\phi_I^{(+)}}{\partial x}\right)^2+
\left(\frac{\partial\phi_I^{(-)}}{\partial x}\right)^2
\right\}
+\frac{i}{4\pi\nu_I}
\frac{\partial\phi_I^{(+)}}{\partial \tau}
\frac{\partial\phi_I^{(-)}}{\partial x}
\right]
\end{equation}
Short-range electron-electron interactions have the form
$V=V_{11}+V_{22}+2V_{12}$
where $V_{II}$'s are intra-channel interactions:
$V_{II}=\pi g_{II}\int dx
\left\{
\left(\rho^{(u)}_I(x)\right)^2+
\left(\rho^{(l)}_I(x)\right)^2
\right\}$
for $I=1,2$ and $V_{12}$ is the inter-channel interaction:
$V_{12}=\pi g_{12}\int dx
\left\{
\rho^{(u)}_1(x)\rho^{(u)}_2(x)+
\rho^{(l)}_1(x)\rho^{(l)}_2(x)
\right\}$.
All $g_{IJ}$'s are positive since we consider the repulsive Coulomb
interaction. The most general form of the edge action for the
multi-channel FQH states has the form
\begin{equation}
S_0 = \sum_{I,J}\int d\tau dx
\left\{
\frac{1}{8\pi}v_{IJ}
\left(
\frac{\partial\phi_I^{(+)}}{\partial x}
\frac{\partial\phi_J^{(+)}}{\partial x}
+
\frac{\partial\phi_I^{(-)}}{\partial x}
\frac{\partial\phi_J^{(-)}}{\partial x}
\right)
+\frac{i}{4\pi}K_{IJ}
\frac{\partial\phi_I^{(+)}}{\partial \tau}
\frac{\partial\phi_J^{(-)}}{\partial x}
\right\}
\end{equation}
In our two-channel model, since two matrices $K$ and $v$ are given by
\begin{equation}
K = \left[
\begin{array}{cc}
1/\nu_1 & 0 \\
0 & 1/\nu_2
\end{array}
\right],\ \ \ 
v = \left[
\begin{array}{cc}
v_1/\nu_1 & g_{12} \\
g_{12} & v_2/\nu_2
\end{array}
\right]
\end{equation}
where $v_I=c_I+\nu_I g_{II}$, the action (2) reduces
to the following special form
\begin{equation}
S_0 = \frac{1}{\beta L}\sum_{\omega,k}\frac{k^2}{8\pi}
\phi^T(-\omega,-k)A(\omega,k)\phi(\omega,k)
\end{equation}
where
$\phi^T = \left(\phi_1^{(+)},\phi_1^{(-)},
\phi_2^{(+)},\phi_2^{(-)}\right)$.
$A(\omega,k)$ is a 4 by 4 matrix which has the form
\begin{equation}
A = \left[
\begin{array}{rr}
A_1 & g_{12}I \\
g_{12}I & A_2
\end{array}
\right]
\end{equation}
where {\it I} is the 2$\times$2 identity matrix, and
\begin{equation}
A_J = \left[
\begin{array}{rr}
a_J & -ib_J \\
-ib_J & a_J
\end{array}
\right]
\end{equation}
with $a_J=v_J/\nu_J, b_J=\omega/\nu_Jk$ for $J =1,2$.
When we introduce backward scatterings between the edges
by making a point contact, the system is described as the TL model
with a scattering potential at $x=0$.
Since here we do not need the detailed structure of the
scattering potential $U[\phi(\tau,x=0)]$,
we leave the analysis of it to the next section.
All we need here is that the total action has the form
$S = S_0 + \int d\tau U[\phi(\tau,x=0)]$.
The partition function $Z = \int{\cal D}\phi e^{-S}$ of the system
can be written in terms of the field
$\theta(\tau) = \phi(\tau,x=0)$ as
\begin{equation}
Z = \int{\cal D}\theta\int{\cal D}\lambda\int{\cal D}\phi
e^{-S_0[\theta,\lambda,\phi]-\int d\tau U[\theta(\tau)]}
\end{equation}
where
\begin{equation}
S_0[\theta, \lambda, \phi] = S_0 + i\int d\tau 
\lambda^T(\tau)\left(\theta(\tau) - \phi(\tau,x=0)\right)
\end{equation}
with
$\lambda^T = \left(\lambda_1^{(+)},\lambda_1^{(-)},
\lambda_2^{(+)},\lambda_2^{(-)}\right)$,
$\theta^T = \left(\theta_1^{(+)},\theta_1^{(-)},
\theta_2^{(+)},\theta_2^{(-)}\right)$
by introducing the auxiliary field $\lambda(\tau)$
so that we may obtain
$\theta(\tau) = \phi(\tau,x=0)$
after integrating out the $\lambda(\tau)$'s.
We perform successive integrations as
$\int{\cal D}\lambda\int{\cal D}\phi
e^{-S_0[\theta,\lambda,\phi]}
= \int{\cal D}\lambda
e^{-S_0[\theta,\lambda]}
= e^{-S_0[\theta]}$
where $S_0[\theta,\lambda]$ has the form
\begin{equation}
S_0[\theta,\lambda] = \frac{1}{\beta}\sum_{\omega}
\left(
\lambda^T(-\omega)B(\omega)\lambda(\omega)
+ i\lambda^T(-\omega)\theta(\omega)
\right)
\end{equation}
with
\begin{equation}
B(\omega)
= \int_{-\infty}^{\infty}\frac{dk}{k^2}A^{-1}(\omega,k)
= \oint\frac{\tilde{A}(\omega,k)dk}{k^2\det A(\omega,k)}
\end{equation}
where $\tilde{A}(\omega,k)$ is the adjugate matrix of
$A(\omega ,k)$.
The matrix $B(\omega)$ can be explicitly evaluated.
The determinant of $A(\omega,k)$ is given by
$\det A = g^4-2g^2(a_1a_2-b_1b_2)
+(a_1^2+b_1^2)(a_2^2+b_2^2)$
where
$g=\sqrt{(v_1-v_2)^2+4\xi}$ with $\xi =\nu_1\nu_2g_{12}^2$.
It is equivalent to
\begin{equation}
\nu _1^2\nu _2^2k^4\det A = \eta^2(k^2+k_+^2)(k^2+k_-^2)
\end{equation}
where $\eta =v_1v_2-\xi$.
$k_\pm$ are positive constants which satisfy
\begin{equation}
k_+^2+k_-^2=(v_1^2+v_2^2+2\xi)\omega^2/\eta^2,\ \ \ 
k_+^2k_-^2=\omega^2/|\eta|.
\end{equation}
It turns out that finite components are
$B_{11}=B_{22}$, $B_{33}=B_{44}$, and
$B_{13}=B_{24}=B_{31}=B_{42}$.
The other components vanish since their integrands have odd powers of k.
We obtain
\begin{equation}
B_{11}=\frac{\pi\nu_1}{\eta^2}
\frac{v_1|\eta|+v_2\eta}{k_++k_-},\ \ \ 
B_{33}=\frac{\pi\nu_2}{\eta^2}
\frac{v_1\eta+v_2|\eta|}{k_++k_-},\ \ \ 
B_{13}=\frac{\pi g\nu_1\nu_2}{\eta^2}
\frac{|\eta|-\eta}{k_++k_-}
\end{equation}
where $p=(v_1-v_2)/g$, $q=2\nu_1\nu_2g_{12}/g$.
Rearranging the fields as
$\lambda^{(\pm)T} = \left(\lambda_1^{(\pm)},\lambda_2^{(\pm)}\right)$,
$\theta^{(\pm)T} = \left(\theta_1^{(\pm)},\theta_2^{(\pm)}\right)$,
we can write the action (9) in the form
$S_0[\theta,\lambda]
=S_0^{(+)}[\theta^{(+)},\lambda^{(+)}]
+S_0^{(-)}[\theta^{(-)},\lambda^{(-)}]$,
where
\begin{equation}
S_0^{(\pm)}[\theta^{(\pm)},\lambda^{(\pm)}]
= \frac{1}{\beta}\sum_{\omega}
\left(
\frac{\pi}{|\omega|}\lambda^{(\pm)T}(-\omega)P
\lambda^{(\pm)}(\omega)
+i\lambda^{\pm T}(-\omega)\theta^{\pm}(\omega)
\right)
\end{equation}
The matrix P is given by
\begin{equation}
P =
\left\{ 
\begin{array}{rr}
\frac{v_1+v_2}{|v_1+v_2|}\left[
\begin{array}{rr}
\nu_1 & 0 \\
0 & \nu_2
\end{array}
\right]
&
(\eta>0)
\\
\\
\left[
\begin{array}{rr}
\nu_1p & q \\
q & -\nu_2p
\end{array}
\right]
&
(\eta<0)
\end{array}
\right.
\end{equation}
Integrating out $\lambda$'s, we finally obtain the effective action
for $\theta$ as 
$S_0[\theta]
=S_0^{(+)}[\theta^{(+)}]+S_0^{(-)}[\theta^{(-)}]$
with
\begin{equation}
S_0^{(\pm)}[\theta^{(\pm)}]
= \frac{1}{4\pi\beta}\sum_{\omega}
|\omega|\theta^{(\pm)T}(-\omega)P^{-1}\theta^{(\pm)}(\omega).
\end{equation}
The partition function of the system now reduces to
$Z = \int{\cal D}\theta e^{-S_0[\theta]-
\int d\tau U[\theta(\tau)]}$.
Note that $\theta^{(+)}$ and $\theta^{(-)}$ are decoupled in $S_0$.

\subsection{Stability condition}
\noindent
In later sections, we will show that the scaling dimensions of
the tunneling amplitudes are universal if both of the channels
propagate in the same direction.
The stability of the system is crucial for this
universality.

The stability of the system requires that the Hamiltonian should
be positive definite. In order to proceed it is convenient to
transform the action into a representation in which both $K$ and $v$
are diagonal. This can be accomplished by performing the following
transformations ~\cite{Kane95,Wen95}:
$\phi = \Lambda\tilde{\phi} = 
\Lambda_1\Lambda_2\Lambda_3\tilde{\phi}$,
where the first step is the diagonalization of the $K$ matrix: 
$\left(\Lambda_1^TK\Lambda_1\right)_{IJ} = \delta_{IJ}/\nu_I$
with $(\Lambda_1^T\Lambda_1)_{IJ} = \delta_{IJ}$.
The second step is given by
$\left(\Lambda_2\right)_{IJ} = \sqrt{|\nu_I|}\delta_{IJ}$ and
$\Lambda_2^T\Lambda_1^TK\Lambda_1\Lambda_2= \Sigma$
where $\Sigma_{IJ}=\sigma_{I}\delta_{IJ}$
with $\sigma_{I}=\nu_I/|\nu_I|$.
In the final step we choose as
$\Lambda_3^T\Sigma\Lambda_3 = \Sigma$ and
$\Lambda_3^T\Lambda_2^T\Lambda_1^Tv\Lambda_1\Lambda_2\Lambda_3
= \tilde{v}$
with $\tilde{v}=\tilde{v}_I\delta_{IJ}$.
In the $\tilde{\phi}$-representation the action (2) has the form
in which all the channels are decoupled,
\begin{equation}
S_0 = \sum_{I}\int d\tau dx
\left[
\frac{1}{8\pi}\tilde{v}_I
\left\{
\left(\frac{\partial\phi_I^{(+)}}{\partial x}\right)^2+
\left(\frac{\partial\phi_I^{(-)}}{\partial x}\right)^2
\right\}
+\frac{i}{4\pi}\sigma_I
\frac{\partial\phi_I^{(+)}}{\partial \tau}
\frac{\partial\phi_J^{(-)}}{\partial x}
\right].
\end{equation}
The situation is similar for the Hamiltonian
\begin{eqnarray}
H &=& H^{(u)}+H^{(l)} =
\frac{1}{8\pi}\int dx\sum_{I,J}v_{IJ}
\left(
\rho^{(u)}_I(x)\rho^{(u)}_J(x)+
\rho^{(l)}_I(x)\rho^{(l)}_J(x)
\right)
\nonumber\\
&=& \frac{1}{8\pi}\int dx\sum_{I}\tilde{v}_I
\left\{
\left(\tilde{\rho}^{(u)}_I(x)\right)^2+
\left(\tilde{\rho}^{(l)}_I(x)\right)^2
\right\}
\end{eqnarray}
where 
$\rho_I^{(u,l)}(x) = J_{0I}^{(u,l)}(x) = 
\partial_1\phi_I^{(u,l)}(x)/2\pi$,
and
$\tilde{\rho}_I^{(u,l)}(x) = \tilde{J}_{0I}^{(u,l)}(x) = 
\partial_1\tilde{\phi}_I^{(u,l)}(x)/2\pi$
satisfy the following commutation relations,
\begin{equation}
[\rho_I^{(u,l)}(k), \rho_{J}^{(u,l)}(-k)]=
\pm\frac{k}{2\pi}\left(K^{-1}\right)_{IJ},\ \ \ 
[\tilde{\rho}_I^{(u,l)}(k), \tilde{\rho}_{J}^{(u,l)}(-k)]=
\pm\frac{k}{2\pi}\sigma_I\delta_{IJ}
\end{equation}
The stability of the system requires that all $\tilde{v}_I$'s should
be positive. In our two-channel model, $K$ and $v$ are given in (3).
Since $K$ has been already been diagonalized, $\Lambda_1$ is chosen
to be an identity. The choice of $\Lambda_2$ is trivial.
Finally we must choose $\Lambda_3$ such that 
$\Lambda_3^T\Sigma\Lambda_3 = \Sigma$.
Hence it is enough to study the following two cases ~\cite{Wen92}.

\noindent
(i)$\nu_1\nu_2>0$

\noindent
In this case, the matrix $\Sigma$ is the identity except its sign,
$\Sigma=\sigma I$ where $\sigma=\sigma_1=\sigma_2$.
The matrix $\Lambda_3$ can be chosen as 
$\Lambda_3 = \left[
\begin{array}{rr}
\cos\theta & \sin\theta \\
-\sin\theta & \cos\theta
\end{array}
\right]$
where $\theta$ must satisfy
$\tan 2\theta=-2\sigma\sqrt{\xi}/(v_1-v_2)$.
It can be shown that $\tilde{v}_1$ and $\tilde{v}_2$
are two different solutions of the quadratic equation
$\tilde{v}^2-\sigma(v_1+v_2)\tilde{v}+\eta = 0$.
Note that the discriminant $D=g^2$ is always positive,
which ensure the real solutions for $v_1$ and $v_2$.
Hence the stability condition reduces to
$\sigma(v_1+v_2)>0$ and $\eta>0$.
In this case, the matrix $P$ given in (11) becomes diagonal
and has only positive universal components.
Recall that $\eta =v_1v_2-\xi$ with $\xi =\nu_1\nu_2g_{12}^2$.
In the present case since both $\xi$ and $\eta$ are positive,
$v_1v_2>0$. Hence both channels propagate in the same direction.

\noindent
(ii)$\nu_1\nu_2<0$

\noindent
In this case, the matrix
$\Lambda_3$ which satisfies
$\Lambda_3^T\Sigma\Lambda_3 = \Sigma$ has the form
$\Lambda_3 = \left[
\begin{array}{rr}
\cosh\theta & \sinh\theta \\
\sinh\theta & \cosh\theta
\end{array}
\right]$
where
$\tanh 2\theta=-2\sigma_1\sqrt{-\xi}/(v_1-v_2)$.
It is shown that
$\tilde{v}_1$ and $\tilde{v}_2$
satisfy the following quadratic equation respectively:
$\tilde{v}_{1,2}^2+\sigma_{1,2}(v_1+v_2)\tilde{v}_{1,2}+\eta = 0$.
The requirement of stability is that both
$\tilde{v}_1$ and $\tilde{v}_2$ should have a positive solution,
which is equivalent to $\eta<0$.
The matrix $P$ is non-universal and has off-diagonal components.
In the present case since both $\xi$ and $\eta$ are negative,
$v_1v_2<0$. Hence both channels propagate
in the opposite direction.

\section{Quasiparticle tunneling picture}
\noindent
In this section, we first consider the limit where the scattering
potential is very weak, i.e.
we start the RG analysis from the situation where the upper and
lower branches are separated by the bulk FQH liquid.
Quasiparticle tunnelings are possible only near the 
point contact since only there not only
upper and lower branches approach 
but also the translational symmetry breaks down.

Performing the RG transformation, we calculate exactly the scaling
dimensions of the scattering amplitudes.
Next we consider the opposite limit where the scattering potential
is very strong.
In this limit, the electron transport can be viewed as the tunneling
of the phase $\theta$
from a potential minimum to an adjacent minimum.
This process corresponds to an instanton or an anti-instanton.
By the duality mapping we transform the original model to an analogous model
in the weak potential limit and
RG analysis of the tunneling amplitude is given.

It will be shown that the scaling dimensions are universal when
$\nu_1\nu_2>0$, while it depends on the interaction when $\nu_1\nu_2<0$.
In case (i) most of the processes are relevant, but
for the (1,$-$1/3) state only $\alpha_{22}$ is positive,
which results in a different situation
for the tunneling at low temperature.

\subsection{Quasiparticle tunneling at a point contact}
\noindent
Now we analyze in detail the backward scatterings at a
point contact.
The scattering potential at $x=0$ has the form
$U=U_{11}+U_{22}+U_{12}$
where $U_{II}$ corresponds to the intra-channel quasiparticle
tunneling process:
\begin{equation}
U_{II} = -u_{II}\cos\theta_I^{(+)}(\tau) \propto
\left(\Psi_I^{(u)\dagger}\Psi_I^{(l)}\right)^{\nu_I}+h.c.
\end{equation}
where $\Psi_I^{(u,l)}$ is the electron operator
on the upper (lower) edge. In terms of the bose field
$\theta_I^{(u,l)}$ the electron operator
can be written as
\begin{equation}
\Psi^{(u,l)}_I(\tau)=
\sqrt{\bar{\rho}}\exp\left(\pm i\theta^{(u,l)}_I(\tau)/\nu_I\right)
\end{equation}
A similar formula is given in (A5) in the appendix.
The other term, which will be proven to be important
for our following discussions,
describes inter-channel scatterings. For $\nu_1\nu_2>0$,
it has the form
$U_{12}=U_{12}^{ul}+U_{12}^{lu}$ where
\begin{eqnarray}
U_{12}^{ul} &=& -u_{12}^{ul}\cos
\frac{\theta_+^{(+)}(\tau)+\theta_-^{(-)}(\tau)}{2}
\propto
\left(\Psi_1^{(u)}\Psi_2^{(l)\dagger}\right)^{\nu_1}+h.c.,
\nonumber\\
U_{12}^{lu} &=& -u_{12}^{lu}\cos\
\frac{\theta_+^{(+)}(\tau)-\theta_-^{(-)}(\tau)}{2}
\propto
\left(\Psi_1^{(l)}\Psi_2^{(u)\dagger}\right)^{\nu_1}+h.c.
\end{eqnarray}
For $\nu_1\nu_2<0$, $U_{12}=U_{12}^{uu}+U_{12}^{ll}$
where
\begin{eqnarray}
U_{12}^{uu} &=& -u_{12}^{uu}\cos
\frac{\theta_+^{(+)}(\tau)+\theta_+^{(-)}(\tau)}{2}
\propto
\left(\Psi_1^{(u)}\Psi_2^{(u)}\right)^{\nu_1}+h.c.,
\nonumber\\
U_{12}^{ll} &=& -u_{12}^{ll}\cos\
\frac{\theta_+^{(+)}(\tau)-\theta_+^{(-)}(\tau)}{2}
\propto
\left(\Psi_1^{(l)}\Psi_2^{(l)}\right)^{\nu_1}+h.c.
\end{eqnarray}
Here
$\theta_{\pm}^{(\pm)}=
\theta_1^{(\pm)}\pm\rho\theta_2^{(\pm)}$ and
$\rho=\nu_1/\nu_2$,
and we assumed that $|\nu_1|>|\nu_2|$ with
$\rho$ being an odd integer.
$U_{12}$ corresponds to inter-branch
(intra-branch) tunnelings for $\nu_1\nu_2>0$ ($\nu_1\nu_2<0$).
In the following we also assume $u_{12}^{ul}=u_{12}^{lu}=u_{12}$
for $\nu_1\nu_2>0$ and $u_{12}^{uu}=u_{12}^{ll}=u_{12}$
for $\nu_1\nu_2<0$. Hence for $\nu_1\nu_2>0$
\begin{equation}
U_{12} = -2u_{12}\cos\frac{\theta_+^{(+)}(\tau)}{2}
\cos\frac{\theta_-^{(-)}(\tau)}{2}
\end{equation}
and for $\nu_1\nu_2<0$
\begin{equation}
U_{12} = -2u_{12}\cos\frac{\theta_+^{(+)}(\tau)}{2}
\cos\frac{\theta_+^{(-)}(\tau)}{2}.
\end{equation}
We assume that all $u_{IJ}$'s are positive but the result
does not depend on the signs of $u_{IJ}$'s.

\subsection{Weak potential limit}
\noindent
We first consider the limit where the scattering potential is
very weak.
We calculate the scaling dimensions of the scattering amplitudes
by recursively integrating out the high-frequency modes 
~\cite{Kane92}~\cite{Furusaki93}~\cite{Fisher85}.
At zero temperature, the partition function can be written as
$Z = \int{\cal D}\theta_{\Lambda}
e^{-S_0[\theta_{\Lambda}]-
\int d\tau U_{\Lambda}[\theta_{\Lambda}(\tau)]}$.
Here we introduced a cutoff $\Lambda$ such that
\begin{equation}
\theta_{\Lambda}(\omega) =
\left\{
\begin{array}{ll}
\theta(\omega) & (|\omega|<\Lambda)\\
0 & (\Lambda<|\omega|)
\end{array}
\right.
\end{equation}
The unperturbed action $S_0[\theta_{\Lambda}]$ has the form
$S_0[\theta_{\Lambda}]
=S_0^{(+)}[\theta_{\Lambda}^{(+)}]
+S_0^{(-)}[\theta_{\Lambda}^{(-)}]$
where
\begin{equation}
S_0^{(\pm)}[\theta_{\Lambda}^{(\pm)}]
= \frac{1}{4\pi\beta}\sum_{\omega}
|\omega|\theta_{\Lambda}^{(\pm)T}(-\omega)
P^{-1}\theta_{\Lambda}^{(\pm)}(\omega)
\end{equation}
We divide the field $\theta_{\Lambda}(\tau)$ into slow and fast
modes as
$\theta_{\Lambda}(\tau)=\theta_{\mu}(\tau)+\theta_{fast}(\tau)$
such that
\begin{equation}
\theta_{\Lambda}(\omega) = \left\{
\begin{array}{ll}
\theta_{fast}(\omega) & (\mu<|\omega|<\Lambda)\\
\theta_{\mu}(\omega) & (|\omega|<\mu)
\end{array}
\right.
\end{equation}
Integrating out the fast modes, we obtain an effective action for
the slow modes $\theta_{\mu}$:
$Z = \int{\cal D}\theta_{\mu}
e^{-S_0[\theta_{\mu}]-\int d\tau U(\mu)[\theta_{\mu}(\tau)]}$.
Here we keep only the first order terms
with respect to the scattering
potential U since they are most relevant.
We define the scaling dimension $\alpha_{IJ}$ as
\begin{equation}
\frac{u_{IJ}(\mu)}{\mu} =
\left(\frac{\mu}{\Lambda}\right)^{\alpha_{IJ}}
\frac{u_{IJ}(\Lambda)}{\Lambda}
\end{equation}
for $I,J=1,2$.
These are determined by the equations
$u_{II}(\mu) =
e^{-\frac{1}{2}G_{II}(0)}u_{II}(\Lambda)$
for $I=1,2$, and
$u_{12}(\mu) =
e^{-{1\over 4}
\left(G_{11}(0)+\rho^2G_{22}(0)-2\rho G_{12}(0)
\right)}u_{12}(\Lambda)$.
$G_{IJ}$'s are the correlation functions of the fast modes:
\begin{equation}
G_{IJ}(\tau) = \langle
\theta_{I fast}^{\pm}(\tau)
\theta_{J fast}^{\pm}(0)
\rangle = P_{IJ}\int_{|\omega|<\Lambda}d\omega
\frac{e^{-i\omega\tau}}{|\omega|}W\left({\mu\over\omega}\right)
\end{equation}
where $W(x)$ is a smoothing function with $W(x) \rightarrow 0$
for $|x| \ll 1$ and $W(x) \approx 1$ for $x \gg 1$ ~\cite{Fisher85}.
$\alpha_{IJ}$'s are universal only when $\nu_1\nu_2>0$
corresponding to the form of matrix $P$ given in (15).
The tunneling conductance scales at a ``high'' temperature as
\begin{equation}
G(T)=G(\infty)-cst.\times T^{2\min[\alpha_{IJ}]}
\end{equation}
where $G(\infty)$ is the quantized Hall conductance.
The exponent $\alpha_{IJ}$'s are listed in Table 1, and explicit values
are presented for $\nu=2/5$ and 2/3 cases in Table 2.

\subsection{Strong potential limit}
\noindent
Here we consider the limit where the scattering potential is
very strong.
In this limit, the electron transport can be viewed as the tunneling
from a potential minimum to its adjacent one, and tunneling matrix
elements, i.e. instanton fugacities are the expansion parameters.
Using the dilute instanton gas approximation (DIGA), we transform
the original model to its dual theory in the weak potential limit
and apply to it the RG analysis in the preceding subsection.
In determining the minima of the scattering potential
$U[\theta]$ at $x=0$, only
$\theta_1=\theta_1^{(+)}$, $\theta_2=\theta_2^{(+)}$, and
$\theta_3=\theta_+^{(-)}$ 
($\theta_3=\theta_-^{(-)}$)
for $\nu_1\nu_2>0$ ($\nu_1\nu_2<0$)
are independent.
After integrating out the other degree of freedom,
the partition function has the form
$Z = \int{\cal D}\theta_1\int{\cal D}
\theta_2\int{\cal D}\theta_3
e^{-S_0[\theta_1,\theta_2,\theta_3]-
\int d\tau U[\theta_1,\theta_2,\theta_3]}$
where
\begin{equation}
S_0[\theta_1,\theta_2,\theta_3]
= \frac{1}{4\pi\beta}\sum_{\omega}
|\omega|\theta^{T}(-\omega)Q\theta(\omega)
\end{equation}
with
$\theta^T=(\theta_1,\theta_2,\theta_3)$ and
Q is a 3 by 3 matrix which has the form
\begin{equation}
Q = \left(
\begin{array}{cc}
P^{-1} & \vec{0} \\
\vec{0}^T & 1/\nu_3
\end{array}
\right)
\end{equation}
where $\nu_3$ is defined as
\begin{equation}
\nu_3 = \left\{
\begin{array}{ll}
|\nu_1|+\rho^2|\nu_2|        & (\eta>0)\\
(\nu_1-\rho^2\nu_2)p+2\rho q & (\eta<0)
\end{array}
\right.
\end{equation}
The scattering potential (22-25) now has the form
$U=U_{1}+U_{2}+U_{3}$
where
\begin{eqnarray}
U_{1} &=& -u_{1}\cos\theta_1(\tau),
\nonumber\\
U_{2} &=& -u_{2}\cos\theta_2(\tau),
\nonumber\\
U_{3} &=& -u_{3}\cos\frac{\theta_1(\tau)+\rho\theta_2(\tau)}{2}
\cos\frac{\theta_3}{2}.
\end{eqnarray}
Hence the scattering potential has minima at
$(\theta_1,\theta_2,\theta_3)=2\pi(l,m,n)$
where $l,m,n$ are integers which satisfy the condition that
$l+m$ is (a)even for even n and (b) odd for odd n.
It follows that the minima of the potential form a
double-layer lattice in the three-dimensional
$\theta_1\theta_2\theta_3$-space.
The typical layers of that lattice are the equi-$\theta_3$ planes
with (a) $\theta_3=4k\pi$ and (b) $\theta_3=2(2k+1)\pi$.

Using the DIGA we show below that the partition function
in the strong  potential limit is mapped to an analogous
one in the weak potential limit~\cite{Schmid83}.
In contrast to the single-channel case, it turns out that
the dual theory is not identical to the original one.

In the DIGA the field $\theta_I(\tau)$ can be written as
$\theta_I(\tau)=\sum_{j=1}^{9}C_{Ij}X_j(\tau)$
where $C_{Ij}$'s are listed in Table 3.
$X_j$'s are linear combinations of instantons ($e_{jk}=1$)
and anti-instantons ($e_{jk}=1$):
$X_j(\tau)=\sum_{k=1}^{n_j}e_{jk}X(\tau-\tau_{jk})$
where $e_{jk}=1$ and $\tau_{jk}$ are respectively
the topological charge and the central imaginary time of
the k-th instanton in the j-th species.
$X(\tau)$ is a one-instanton solution at $\tau=0$
the width of which should be neglected compared to
$\beta$ in the DIGA.
The instanton or anti-instanton in $X_j(\tau)$
which we call those of the
$j$-th species corresponds to a tunneling process from
$(\theta_1,\theta_2,\theta_3)$ to
$(\theta_1,\theta_2,\theta_3)+2\pi e_{jk}(C_{1j},C_{2j},C_{3j})$.
Here we took into account only the nine species listed
in Table 3
since they are most likely to be relevant.

Tunneling of an instanton or an anti-instanton
along the $\theta_1$-axis, $\theta_2$-axis, and
$\theta_3$-axis can be interpreted physically
as a pulse of 1-channel current,
2-channel current and the total charge density fluctuation respectively.
The bosonization of current and density is given for example
in the appendix, where they are expressed as
$J_{I}=\partial_0 \theta_I^{(+)}/2\pi$ and
$\rho_{I}=\partial_0 \theta_I^{(-)}/2\pi$.

We may write the Fourier transform of $\theta_I(\tau)$ as
\begin{equation}
\theta_I(\omega)=\frac{2\pi i}{\omega}
\sum_{j=1}^{9}C_{Ij}\sum_{k=1}^{n_j}e_{jk}e^{i\omega\tau_{jk}}
\end{equation}
where we used the approximation
$dX/d\tau\approx 2\pi\delta(\tau)$.

The partition function now has the form
$Z=\sum e^{-S_0[\theta_1,\theta_2,\theta_3]}$
where $\theta_I$'s satisfy (36).
$\sum$ denotes a "summation" over instanton configurations,
or explicitly,
\begin{equation}
\sum = \prod_{j=1}^{9}\sum_{n_j=0}^{\infty}
\frac{y_j^{n_j}}{n_j!}
\sum_{\{e_{jk}\}}\prod_{k=1}^{n_j}\int_0^{\beta}d\tau_{jk}
\end{equation}
where $y_j$ is the instanton fugacity, i.e.
the tunneling matrix elements from
$(\theta_1,\theta_2,\theta_3)=(0,0,0)$ to
$2\pi e_{jk}(C_{1j},C_{2j},C_{3j})$.
We consider the Gaussian integral over the"dual fields":
\begin{equation}
\int{\cal D}\tilde{\theta}
\exp\left(-\frac{1}{4\pi\beta}\sum_{\omega}
|\omega|\tilde{\theta}^{T}(-\omega)Q^{-1}
\tilde{\theta}(\omega)\right)
\end{equation}
where
$\tilde{\theta}^T=
(\tilde{\theta_1},\tilde{\theta_2},\tilde{\theta_3})$
and insert it into the "summation" in the partition function.
Next we perform the shift transformation
$\tilde{\theta}(\omega) \rightarrow \tilde{\theta}(\omega)
+\left(\omega/|\omega|\right)\theta(\omega)$
which cancels the original action
$S_0[\theta_1,\theta_2,\theta_3]$.
We can write the partition function in the form
\begin{equation}
Z=\sum \int{\cal D}\tilde{\theta}
\exp\left[
-\tilde{S}_0[\tilde{\theta}]-
i\sum_{j=1}^{9}\sum_{k=1}^{n_j}e_{jk}
\sum_{I=1}^{3}C_{Ij}\tilde{\theta}_I(\tau_{jk})
\right]
\end{equation}
where
\begin{equation}
\tilde{S}_0[\tilde{\theta}]
= \frac{1}{4\pi\beta}\sum_{\omega}
|\omega|\tilde{\theta}^{T}(-\omega)Q^{-1}\tilde{\theta}(\omega)
\end{equation}
Taking the ``summation'' over instanton configurations,
we can finally write the partition function in terms of
the dual action $\tilde{S}[\tilde{\theta}]$ as
$Z = \int{\cal D}\tilde{\theta}e^{-\tilde{S}[\tilde{\theta}]}$
where
$\tilde{S}[\tilde{\theta}] =
\tilde{S}_0[\tilde{\theta}]+\int d\tau Y[\tilde{\theta}(\tau)]$
with ``the scattering potential''
\begin{equation}
Y[\tilde{\theta}] = \sum_{j=1}^{9}2y_j\cos
\left(
\sum_{I=1}^{3}C_{Ij}\tilde{\theta}_I
\right)
\end{equation}
It turns out that the unperturbed part
$\tilde{S}_0[\tilde{\theta}]$
of the dual action can be identified with the original
$S_0[\theta]$ by the correspondences
$Q^{-1} \leftrightarrow Q$ and
$\tilde{\theta} \leftrightarrow \theta$.
In contrast to the single-channel case, the potential term
$Y[\tilde{\theta}]$ is analogous but not identical to
the original potential $U[\theta]$
in spite of the correspondence
$2y_j \leftrightarrow -t_I$.

It is straight-forward to calculate the scaling dimensions
$\beta_j$ of the tunneling matrix element $y_j$,
which is defined by
\begin{equation}
\frac{y_{j}(\mu)}{\mu} =
\left(\frac{\mu}{\Lambda}\right)^{\beta_{j}}
\frac{y_{j}(\Lambda)}{\Lambda}
\end{equation}
These are calculated using the equations
\begin{eqnarray}
y_{1}(\mu) &=&
e^{-\frac{1}{2}\left(
\tilde{G}_{11}(0)
+\tilde{G}_{22}(0)
+2\tilde{G}_{12}(0)
\right)}
y_{1}(\Lambda),
\nonumber\\
y_{2}(\mu) &=&
e^{-\frac{1}{2}\left(
\tilde{G}_{11}(0)
+\tilde{G}_{22}(0)
-2\tilde{G}_{12}(0)
\right)}
y_{2}(\Lambda),
\nonumber\\
y_{3,4}(\mu) &=&
e^{-\frac{1}{2}\left(
\tilde{G}_{22}(0)
+\tilde{G}_{33}(0)
\right)}
y_{3,4}(\Lambda),
\nonumber\\
y_{5,6}(\mu) &=&
e^{-\frac{1}{2}\left(
\tilde{G}_{33}(0)
+\tilde{G}_{11}(0)
\right)}
y_{5,6}(\Lambda),
\nonumber\\
y_{7,8,9}(\mu) &=&
e^{-2\tilde{G}_{11,22,33}(0)}
y_{7,8,9}(\Lambda),
\end{eqnarray}
where $\tilde{G}_{IJ}$'s are the correlation functions of
the fast modes:
\begin{equation}
\tilde{G}_{IJ}(\tau) = \langle
\tilde{\theta}_{I fast}^{\pm}(\tau)
\tilde{\theta}_{J fast}^{\pm}(0)
\rangle = Q_{IJ}\int_{|\omega|<\Lambda}d\omega
\frac{e^{-i\omega\tau}}{|\omega|}W\left({\mu\over\omega}\right).
\end{equation}
The results are listed in Table 4.
The explicit values of $\beta_j$ for some specific edge
constructions are shown in Table 5.
In the following we give a detailed discussion on the temperature
dependence of the tunneling conductance in the dilute
instanton gas (DIG) picture.

\subsubsection{DIG picture for $\nu_1\nu_2>0$}
First we consider the case where both of the edge channels
propagate in the same direction.
In Sec. II we have shown that 
$\alpha_{IJ}$'s are universal only for $\nu_1\nu_2>0$,
which is also the case with $\beta_{j}$'s (Table 4).

Table 5 of $\beta_{j}$'s tells us that
all instanton fugacities scale to zero when $\nu_1\nu_2>0$ (i).
In this case we can determine the temperature
dependence of the tunneling conductance using DIG picture:
$G(T) \propto T^{\delta_{DIG}}$
where $\delta_{DIG}=2\min[\beta_j]$.
It should also be noted that the inter-channel tunnelings
play a crucial role in determining $\delta_{DIG}$.
In Table 6 the explicit values of $\delta_{DIG}$ is shown,
for comparison,
in the presence and in the absence
of inter-channel tunnelings.

\subsubsection{DIG picture for $(\nu_1,\nu_2)=(1,-1/3)$} 
Here we consider the (1,$-$1/3) state
which belongs to case (ii): $\nu_1\nu_2<0$.
As was shown in Table 4 the exponents depend on the
interaction between the channels, and not all of the
instanton fugacities scale to zero, i.e. 
the one ($y_9$) which corresponds to $\theta_3=\theta_{-}^{(-)}$
scales to a larger value.
In Sec. II we have shown that the scattering potential $U_{12}$
which suppress the fluctuation of $\theta_3=\theta_{-}^{(-)}$ is
irrelevant.
Hence at both strong and weak potential limit $\theta^{(-)}$
is not pinned.

In contrast to the neutral mode $\theta_3=\theta_{-}^{(-)}$
all $y_j$'s which describe the instanton tunneling
in the $\theta_1^{(+)}$ or $\theta_2^{(+)}$ direction
scale to zero, i.e. both $\theta_1^{(+)}$ and $\theta_2^{(+)}$
are fixed.
However it is shown in Sec. II that only
$\theta_2^{(+)}$ is pinned at weak potential limit.
These two pictures in the strong and weak potential limits
are not smoothly connected.
They can be understood by considering a RG flow diagram
of the type shown in Fig. 1,
which was constructed in the following way.

The inter-channel tunnelings are irrelevant at both of the
strong and weak potential limit for the (1,$-$1/3) state
and can be neglected.
It is enough to study only the intra-channel tunnelings.
In the following we study three DIG pictures. 

First we consider the situation where only $\theta_1^{(+)}$
is pinned.
We start with the action
\begin{equation}
S=S_0^{(+)}[\theta^{(+)}]
+\int_0^{\beta}d\tau
\left(U_1[\theta_1^{(+)}]+U_2[\theta_2^{(+)}]\right)
\end{equation}
where $S_0^{(+)}$ was given in (16), and
$U_1$ and $ U_2$ were introduced in (35).
The scaling dimension of $u_2$ can be calculated as
\begin{equation}
\frac{u_{2}(\mu)}{\mu} =
\left(\frac{\mu}{\Lambda}\right)^{-\nu_{2}p-1}
\frac{u_{2}(\Lambda)}{\Lambda}
\end{equation}
irrespective of the strength of $u_1$.
The scattering potential $u_2$ is
considered to be relevant
also in the DIG picture.
On the other hand on the axis $u_2=0$
the effective action for $\theta_1^{(+)}$
is obtained by
integrating out $\theta_2^{(+)}$:
\begin {equation}
S={1\over4\pi\beta}\sum_{\omega}|\omega|
\theta_1^{(+)}(-\omega)
{1\over\nu_{1}p}\theta_1^{(+)}(\omega)
-u_1\int_0^{\beta}\cos\theta_1^{(+)}(\tau)
\end{equation}
where we used the relation
$p=(v_1-v_2)/g$, $q=2\nu_1\nu_2g_{12}/g$,
$g=\sqrt{(v_1-v_2)^2+4\xi}$ and $\xi =\nu_1\nu_2g_{12}^2$.
We can apply DIGA to the action (47)
to obtain its dual version 
\begin {equation}
\tilde{S}={1\over4\pi\beta}\sum_{\omega}|\omega|
\tilde{\theta}_1^{(+)}(-\omega)
\nu_{1}p
\tilde{\theta}_1^{(+)}(\omega)
-z_1\int_0^{\beta}\cos
\tilde{\theta}_1^{(+)}(\tau)
\end{equation}
where $z_1$ is the instanton fugacity, the scaling dimension
of which is obtained as
\begin{equation}
\frac{z_{1}(\mu)}{\mu} =
\left(\frac{\mu}{\Lambda}\right)^{1/\nu_{1}p-1}
\frac{z_{1}(\Lambda)}{\Lambda}
\end{equation}
Since $1/\nu_{1}p-1<0$ for the $(1, -1/3)$ state
the instanton fugacity $z_1$ is relevant.
In terms of the original scattering potential $u_1$
it tends to decrease in the large-$u_1$ region.
Recall that $u_1$ is irrelevant in the small-$u_1$
region.
We conclude that both the original and the dual picture
belong to the domain (a) of Fig. 1.

Next we consider the situation where only $\theta_2^{(+)}$
is pinned.
In this case the same procedure gives us
\begin{equation}
\frac{u_{1}(\mu)}{\mu} =
\left(\frac{\mu}{\Lambda}\right)^{\nu_{1}p-1}
\frac{u_{1}(\Lambda)}{\Lambda},\ \ \ 
\frac{z_{2}(\mu)}{\mu} =
\left(\frac{\mu}{\Lambda}\right)^{-1/\nu_{2}p-1}
\frac{z_{2}(\Lambda)}{\Lambda}
\end{equation}
where
$-1/\nu_{2}p-1=3/p-1>0$.
It turns out that the scattering potential
$u_1$ is irrelevant but
$u_2$ tends to increase,
which indicates that $(0,\infty)$ is an atractive fixed point
in the $(u_1,u_2)$-plane.

Finally we consider the case where both $\theta_1^{(+)}$
and $\theta_2^{(+)}$ are pinned.
We start with the action (45)
and perform the duality mapping
to obtain
\begin {equation}
\tilde{S}={1\over4\pi\beta}\sum_{\omega}|\omega|
\tilde{\theta}^{(+)}(-\omega)
P
\tilde{\theta}^{(+)}(\omega)
-\sum_{J=1,2}z_J\int_0^{\beta}\cos
\tilde{\theta}_J^{(+)}(\tau)
\end{equation}
where P is a $2\times 2$ matrix given in (15).
The scaling dimensions of $z_1$ and $z_2$ are
calculated as
\begin{equation}
\frac{z_{1}(\mu)}{\mu} =
\left(\frac{\mu}{\Lambda}\right)^{p/\nu_1-1}
\frac{z_{1}(\Lambda)}{\Lambda},\ \ \ 
\frac{z_{2}(\mu)}{\mu} =
\left(\frac{\mu}{\Lambda}\right)^{-p/\nu_{2}-1}
\frac{z_{2}(\Lambda)}{\Lambda}
\end{equation}
where we used the relation
$\det P=-\nu_1\nu_2$.
It turns out that 
both of the instanton fugacities are irrelevant.
In terms of the original scattering potential,
both $u_1$ and $u_2$ tend to increase,
i.e. $(\infty, \infty)$ is another atractive
fixed point in the $(u_1,u_2)$-plane.
It is evident that this picture is not smoothly connected
to the above two DIG pictures where only one of the scattering
potential is strong.
Therefore we expect that there are two domains in the RG flow diagram
in terms of the original scattering potential
$u_1$ and $u_2$.
$(0,\infty)$ and $(\infty, \infty)$ are the
atractive fixed points in the domain (a) and (b) of Fig. 1,
respectively.

\section{Electron tunneling picture}
\noindent
In the latter half of the preceding section, we consider the limit
of strong scattering potential in terms of duality mapping.
There it was revealed that the tunneling amplitude through the potential
is irrelevant in most of the cases.
Then it would be more appropriate to start with the picture
where the bulk FQH state is divided into left and right condensates
and electrons tunnel between the left and right edges
through a insulating region at the point contact.

First we study the electron tunneling (ET) picture
for $\nu_1\nu_2>0$.
As was pointed out before, the situation is different
for the (1,$-$1/3) state. 
In this case
we must consider two different ET
pictures both of which correspond to two atractive fixed points in the
$(u_1,u_2)$-plane, respectively.
These subjects will be discussed toward the end of this section.

\subsection{Construction of the general electron operator}
\noindent
At a point contact, two channels are close to each other within
the magnetic length, the single-channel electron operator
$\Psi_I^{(L,R)}$ on the left (right) branch is no longer the most
general one.
The general electron operator $\Psi^{(L,R)}$ may contain charge
transfers between the edges of the same branch and takes 
the following form ~\cite{Wen92}
\begin{equation}
\Psi^{(L,R)}(s=0)=\sum_{n=-\infty}^{\infty}
c_n\eta^{(L,R)n}\Psi_1^{(L,R)}(x=0)
\end{equation}
where the elementary charge transfer $\eta^{(L,R)}$
is given by
\begin{equation}
\eta^{(L,R)}
=\left(
\Psi_2^{(L,R)}
\Psi_1^{(L,R)\dagger}
\right)^{\nu_1}
= \exp\left[{\pm i\left(
\phi_1^{(L,R)}(x=0)-\rho\phi_2^{(L,R)}(x=0)
\right)}\right]
\end{equation}
The electron tunneling from the left to right edge
is expressed by the operator
$\Gamma = \gamma
\Psi^{(L)}\Psi^{(R)\dagger}|_{x=0}$.

\subsection{ET picture for $\nu_1\nu_2>0$}
\noindent
First we consider the case where the two channels propagate
in the same direction.
The electron tunneling through a point contact can be described
using the concept of tunneling Hamiltonian ~\cite{Mahan81}
$H_T=\Gamma+\Gamma^{\dagger}$
where field operators on the left and right edges anticommute,
which in the present picture corresponds to the fact
that two edges are those of independent incompressible
fluids.
We consider a nonequilibrium situation where
the chemical potentials on each of the edges are different
by the applied voltage: $eV = \mu_L-\mu_R$.

The tunneling current $I(t)$ from the left to right edge is
given by the average value of the operator $edN_L/dt$
where $N_L$ is the number of particles on the left edge.
Since we concentrate on the linear response of $I(t)$ with
respect to $V$, we can evaluate the average using Kubo formula:
$I(t) = ie\int_{-\infty}^t dt'\langle
[dN_L(t)/dt, H_T(t')]\rangle$
where $dN_L(t)/dt=i\left(\Gamma(t)-\Gamma^{\dagger}(t)\right)$.
In terms of the Fourier transform of the 
retarded Green's function
$X_{ret}(\tau) = -i\theta(\tau)\langle
[\Gamma(\tau),\Gamma^{\dagger}(0)]\rangle$
we can write the current as
$I(t) = 2e\Im [X_{ret}(\omega=-eV)]$.
Hence the scaling dimension $\delta_{ET}$ of the tunneling conductance
in the electron tunneling picture
is equal to that of $|\gamma|^2$ ~\cite{Wen91b,Wen92}:
\begin{equation}
\frac{\gamma(\mu)}{\mu} =
\left(\frac{\mu}{\Lambda}\right)^{\delta_{ET}/2}
\frac{\gamma(\Lambda)}{\Lambda}
\end{equation}
We can determine the scaling dimension of $\gamma$
by applying the RG analysis
to the partition function, which can be written
in terms of $\theta^{(L,R)}(\tau) = \phi^{(L,R)}(\tau,x=0)$ as
\begin{equation}
Z = \int{\cal D}\theta e^{
-S_0^{(L)}[\theta^{(L)}]-S_0^{(R)}[\theta^{(R)}]
-\int d\tau H_T[\theta^{(L)}(\tau),\theta^{(R)}(\tau)]}
\end{equation}
where
\begin{equation}
S_0^{(L,R)}[\theta^{(L,R)}]
= \frac{1}{4\pi\beta}\sum_{\omega}
|\omega|\theta^{(L,R)T}(-\omega)
\left[
\begin{array}{cc}
2/|\nu_1| & 0 \\
0 & 2/|\nu_2|
\end{array}
\right]
\theta^{(L,R)}(\omega).
\end{equation}
with
$\theta^{(L,R)T}=(\theta_1^{(L,R)},\theta_2^{(L,R)})$.
Here we have integrated out the fields except
at the point contact.
In the present picture $\theta^{(L)}$ and $\theta^{(R)}$
are not canonically conjugate but are the independent degrees
of freedom.

Next we introduce a cutoff $\Lambda$ and apply the RG
analysis in the preceding section.
Performing the integral over the fast modes, we obtain
\begin{equation}
\gamma(\mu)=
\gamma(\Lambda)
\prod_{L,R}\max
\left[e^{-{1\over2}\left\{ \left({1/|\nu_1|}-n\right)^2\hat{G}_{11}(0)
+n^2\rho^2\hat{G}_{22}(0)\right\}}\right]
\end{equation}
where $\hat{G}_{IJ}$ is
\begin{equation}
\hat{G}_{IJ}(\tau) = \langle
\theta_{I fast}^{(L,R)}(\tau)
\theta_{J fast}^{(L,R)}(0)
\rangle = {|\nu_I|\over 2}\delta_{IJ}\int_{|\omega|<\Lambda}d\omega
\frac{e^{-i\omega\tau}}{|\omega|}W\left({\mu\over\omega}\right)
\end{equation}
Using $\nu_3 = |\nu_1|+\rho^2|\nu_2|>0$
we can write $\delta_{ET}$ as
\begin{equation}
\delta_{ET} = 2\left({1\over|\nu_1|}
+\nu_3\min\left[\left(n-{1\over\nu_3}\right)^2\right]
-{1\over\nu_3}\right)-2.
\end{equation}
It is shown that the minimum value is obtained when
$n$=1 for $(\nu_1,\nu_2)=(1/3,1/15)$,
$n$=1 or 2 for $(\nu_1,\nu_2)=(1/3,1/3)$
and $n$=2 or 3 for $(\nu_1,\nu_2)=(1/5,1/5)$, respectively.
Hence $\delta_{ET}$'s are evaluated as
\begin{equation}
\delta_{ET}=\left\{
\begin{array}{ll}
4\ \ \  & \mbox{for\ \ \ $(\nu_1,\nu_2)$=(1/3,1/15)}\\
4/3\ \ \  & \mbox{for\ \ \ $(\nu_1,\nu_2)$=(1/3,1/3)}\\
16/5& \mbox{for\ \ \ $(\nu_1,\nu_2)$=(1/5,1/5)}
\end{array}
\right..
\end{equation}

\subsection{ET picture for $(\nu_1,\nu_2)=(1,-1/3)$}
For the (1,$-$1/3) state we must consider two different
ET pictures.
In Sec. III we have shown that there are
two attractive fixed points
in the $(u_{1},u_{2})$-plane,
$(0,\infty)$ and $(\infty,\infty)$.
We can consider the ET picture 
in the vicinity of these points.

In Sec. III we performed the RG analysis near
$(u_1, u_2)=(0,0)$, where we discuss the inter-channel
tunneling process $U_{12}$ as well as $U_{11}$ and $U_{22}$. 
Here we concentrate on the analysis of intra-channel
tunnelings $U_{1}=U_{11}$ and $U_{2}=U_{22}$
since the RG flow of $\theta^{(-)}$ have turned out to be
trivial.

For $(u_1, u_2)=(0,\infty)$
we start with the situation where only the 2-channel is reflected.
In this case we slightly modify the discussion in Sec. III
as
\begin{equation}
U_2\rightarrow\Gamma_2=\gamma_2\cos{\theta_2^{(+)}\over\nu_2}.
\end{equation}
since we consider the electron tunneling from the left to right
branch for the 2-channel.
We obtain
\begin{equation}
{\gamma_2(\mu)\over\mu}=\left({\mu\over\Lambda}\right)^{-p/\nu_2-1}
{\gamma_2(\Lambda)\over\Lambda}
\end{equation}
where
$-p/\nu_2-1=3p-1\rightarrow 2$ as $g_{12}\rightarrow 0$.
We confirm that $(0,\infty)$ is a stable fixed point.
Recall that in the DIG picture
the scaling dimension of instanton fugacity $z_1$ in the
present region is given by
$-1/\nu_2 p-1=3/p-1$
which is smaller than $-p/\nu_2-1=3p-1$ since $p>1$.
We predict the existence of a crossover between the two
pictures.

For $(u_1, u_2)=(\infty,\infty)$
we start the RG analysis with the picture where
both of the channels are reflected:
$U_1\rightarrow\Gamma_1=\gamma_1\cos\left(\theta_1^{(+)}/ \nu_1\right)$,
$U_2\rightarrow\Gamma_2=\gamma_2\cos\left(\theta_2^{(+)}/ \nu_2\right)$.
We obtain
$\gamma_1(\mu)/\mu=\left(\mu/\Lambda\right)^{p/ \nu_1-1}
\gamma_1(\Lambda)/\Lambda$,
$\gamma_2(\mu)/\mu=\left(\mu/\Lambda\right)^{-p/ \nu_2-1}
\gamma_2(\Lambda)/\Lambda$
where
$p/ \nu_1-1=p-1>0$ and $-p/ \nu_2-1=3p-1\rightarrow 2$
as $g_{12}\rightarrow 0$.
It is confirmed that the point $(\infty,\infty)$ is another stable fixed point.
In this case the scaling dimensions of the tunneling amplitudes are
precisely equal to those of the instanton fugacities in the DIG
picture.
Hence we predict no crossover between the two pictures.

The whole $(u_1,u_2)$-plane is divided into two domains
separated by a boundary (the dotted line in Fig.1), each of which
has the attractive fixed point
$(0,\infty)$ and $(\infty,\infty)$, respectively.

\section{Discussion and Conclusions}
\noindent
In this paper we have shown that

\noindent
(a) Scaling dimensions are universal only when both
channels propagate in the same direction, which is
guaranteed by requiring the stability of the
system.

\noindent
(b) In the presence of inter-channel tunnelings,
the quasiparticle tunneling picture and the electron
tunneling picture give different temperature dependence
of the tunneling conductance and there exists a crossover
between the two pictures.

\noindent
(c) For the (1, $-$1/3) state the phase diagram is divided
into two domains giving different temperature dependence of the
conductance.

\vspace{1cm}
Finally we explicitly write down our
prediction for the temperature dependence of the tunneling
conductance through a point contact in the presence of interactions.

First we consider the case where $\nu_1\nu_2>0$ (i).
In the case of $(\nu_1,\nu_2)=(1/5,1/5)$
we can read
$\min[\alpha_{IJ}]=-4/5$, $\delta_{DIG}=13$ and
$\delta_{ET}=16/5$
from the tables of exponents. Hence
\begin{equation}
G(T)
\left\{
\begin{array}{ll}
\propto T^{16/5} & \mbox{(at low temperature)}\\
\propto T^{13}  & \mbox{(at intermediate temperature)}\\
={2 \over 5}{e^2 \over h}-cst.\times T^{-8/5} & \mbox{(at ``high'' temperature)}
\end{array}
\right.
\end{equation}
Similarly we obtain for the (1/3,1/15) state
\begin{equation}
G(T)
\left\{
\begin{array}{ll}
\propto T^{4} & \mbox{(at low temperature)}\\
\propto T^{5}  & \mbox{(at intermediate temperature)}\\
={2 \over 5}{e^2 \over h}-cst.\times T^{-28/15} & \mbox{(at ``high'' temperature)\
}
\end{array}
\right.
\end{equation}
and for the (1/3, 1/3) state
\begin{equation}
G(T)
\left\{
\begin{array}{ll}
\propto T^{4/3} & \mbox{(at low temperature)}\\
\propto T^{7}  & \mbox{(at intermediate temperature)}\\
={2 \over 3}{e^2 \over h}-cst.\times T^{-4/3}
& \mbox{(at ``high'' temperature)}
\end{array}
\right.
\end{equation}

\vspace{1cm}
Next we consider the (1, $-$1/3) state which belongs
to case (ii): $\nu_1\nu_2<0$.
As was pointed out in the last section,
our prediction is different
for that state.
There is an ambiguity with the quantized value of the Hall
conductance at ``high'' temperature
when two channels propagate in the opposite
directions.
In this case a generalized
Landauer formula~\cite{Beenakker90,Buttiker86,Buttiker88}
gives different quantized values for the
two-terminal conductance $G_2$ and the four-terminal
Hall conductance $G_H$,
which are both different from
$\nu {e^2 \over h}$~\cite{Kane94,Kane95}.
For $(g_+,g_-)=(\nu_1,|\nu_2|)=(1,1/3)$
they are calculated respectively as~\cite{Kane94,Kane95}
\begin{equation}
G_2={e^2 \over h}(g_{+}+g_{-})={4\over3}{e^2 \over h},\ \ \ 
G_H={e^2 \over h}{g_{+}^2+g_{-}^2 \over g_{+}-g_{-}}={5\over3}{e^2 \over h}.
\end{equation}
However we take the viewpoint that Landauer formula is no longer
valid in this case and the conductance is quantized:
$G(T)\rightarrow{2 \over 3}{e^2 \over h}=\nu{e^2 \over h}$,
which is experimentally observed.

As was studied in Secs. III and IV the RG flow diagram
in the ($u_{1},u_2$)-plane has two stable fixed points,
which means that there are two different behaviors
of the conductance depending on the initial value of $(u_1 ,u_2)$.

First we study the
domain (a) of Fig. 1 where $(u_1,u_2)\rightarrow(0,\infty)$.
If the initial value of $u_1$ is not
so large only the quasihole
$(\nu_2=-1/3)$ channel becomes reflected by the scattering potential
as the temperature is lowered.
The other ($\nu_1=1$ integral quantum Hall)
edge is robust against backward scatterings even at zero
temperature.
At zero temperature the conductance is again
quantized. To obtain the quantized value, Landauer formula for the
two-terminal~\cite{Beenakker90,Buttiker88} and 
four-terminal conductance was applied
~\cite{MacDonald90,Buttiker86}.
Similarly the two-terminal conductance can also be calculated
~\cite{Beenakker90,Buttiker88}.
Both $G_2$ and $G_H$ give the same value ${e^2 \over h}$ as
$T\rightarrow 0$, which is identified with $G(T\rightarrow 0)$.
At low temperature only
$\theta_1^{(+)}$ is relevant for the electron tunneling. 
In this case the Landauer formula is valid
since the single-channel model is free from the complications on the
equilibration between the channels~\cite{Kane94,Oreg95}.

The tunneling conductance behaves as
\begin{equation}
G(T)=
\left\{
\begin{array}{ll}
{e^2 \over h}-cst.\times T^{6p-2} & \mbox{(at low temperature)}\\
{e^2 \over h}-cst.\times T^{6/p-2} & \mbox{(at intermediate temeprature)}\\
{2 \over 3}{e^2 \over h}+cst.\times T^{2p/3-2}
& \mbox{(at ``high'' temperature)}
\end{array}
\right.
\end{equation}
where
$p=(v_1-v_2)/g$, $g=\sqrt{(v_1-v_2)^2+4\xi}$ and $\xi =\nu_1\nu_2g_{12}^2$.
The signs of the temperature dependent terms
reflect the fact that the $\nu_2=-1/3$ edge
is that of a quasihole droplet.
We predict the existence of crossover among the three pictures.

Next we study the
domain (b) of Fig. 1 where $(u_1,u_2)\rightarrow(\infty,\infty)$.
If the initial value of $u_1$ is sufficiently large
both $u_1$ and $u_2$ scale to infinity.
In this case the scaling dimension of
the instanton fugacity $z_1$  in the DIG picture
and that of the electron tunneling amplitude $\gamma_1$ are
identical, i.e. bith give
$p/\nu_1-1=p-1>0$.
The tunneling conductance behaves at low temperature as
\begin{equation}
G(T)\propto T^{2(p-1)}
\end{equation}
where $2(p-1)$ is a positive exponent if $g_{12}$ is finite but
goes to zero as $g_{12}\rightarrow 0$.
We expect that the exponent is small and there is no
crossover between the ET and the DIG picture.
In this case we also predict that even at ``high'' temperature
the tunneling conductance is not quantized since this domain is
not smoothly connected to the origin in the $(u_1,u_2)$-plane.

\vspace{1cm}
In summary we have shown that the structure of the
hierarchical edge modes gives a variety of behaviors of the
conductance through a point contact.
This may give a clue to identify the edge structure in
terms of the experiment.

\acknowledgements
We are grateful to A. Furusaki for useful discussions.
This work was supported by a Grand-in-Aid
for Scientific Research No. 04240103 from the
Ministry of Education, Science and Culture of Japan.

%\vfil\eject
\appendix
\section{Bosonization of the edge mode}
\noindent
We start with the Lagrangian density
\begin{eqnarray}
{\cal L}
&=& \frac{\nu}{4\pi}
\epsilon_{\mu\nu\lambda}
a_{\mu}\partial_{\nu}a_{\lambda}+
\Phi^{\dagger}\left\{i\partial_0-e(A_0+a_0)\right\}\Phi
\nonumber\\
&-& \frac{1}{2m}|\left\{-i\nabla+e(\vec{A}+\vec{a})\right\}\Phi|^2
-V(|\Phi|)
\end{eqnarray}
where $\Phi$ is a boson field and $a_{\mu}$ is a Chern-Simon
gauge field. 
We assume that the interaction between the
bosons is short-ranged:
$V(|\Phi|)=\mu_2 |\Phi|^2 +\mu_4 |\Phi|^4$
where $\mu_2$ is a chemical potential and
$\mu_4$ comes from the Coulomb interaction.
It is clear that the action is minimized
by the trivial constant solution $\Phi=\sqrt{\bar{\rho}}$ with
$\bar{\rho}=-\mu_4/2\mu_2$ and $A+a=\bar{A}+\bar{a}=0$.
Using the equation of motion for $a_{\mu}$ and integrating over
the density fluctuation $\delta\rho=\rho-\bar{\rho}$,
we can write (A1) as
\begin{equation}
{\cal L} = \frac{\nu}{4\pi}\epsilon_{\mu\nu\lambda}
\delta a_{\mu}\partial_{\nu}\delta a_{\lambda}+
{1\over 4\mu_4}\left\{
\partial_0\theta+e(\delta A_0+\delta a_0)
\right\}^2
-\frac{\bar{\rho}}{2m}\left\{
\nabla\theta+e(\delta\vec{A}+\delta\vec{a})
\right\}^2
\end{equation}
where $A=\bar{A}+\delta A$ and $a=\bar{a}+\delta a$.
The single-valued part $\theta_s$ of the phase fluctuation
$\theta$ is the Bogoliubov mode of the bose condensate
whose sound velocity is given by $c^2=2\mu_4\bar{\rho}/m$,
while the multi-valued part $\theta_V$ is the massive
vortex excitation which, we assume, is absent in the
following discussions.
Introducing the Stratonovich-Hubbard variable
$J_{\mu}=(J_0,\vec{J})$,
we can write (A2) in the form
\begin{equation}
{\cal L} = \frac{\nu}{4\pi}\epsilon_{\mu\nu\lambda}
\delta a_{\mu}\partial_{\nu}\delta a_{\lambda}-
\frac{m}{2\bar{\rho}}\left(c^2J_0^2-\vec{J}^2\right)-
J_{\mu}\left\{
\partial_{\mu}\theta_s+e(\delta A_{\mu}+\delta a_{\mu})
\right\}
\end{equation}
where $J_{\mu}$ is the fluctuation of the boson current from
$(\bar{\rho},\vec{0})$.
Integration over $\theta_s$ gives the continuity equation
for the boson current
$\partial_{\mu}J_{\mu}=0$ which is
automatically satisfied by writing $J_{\mu}$ in terms of
the auxiliary gauge field $b_{\mu}$ as
$J_{\mu} =
\epsilon_{\mu\nu\lambda}\partial_{\nu}b_{\lambda}/2\pi$.
After integrating out the gauge field fluctuation $\delta a$,
we obtain
\begin{equation}
{\cal L}= \frac{1}{4\pi\nu}
\epsilon_{\mu\nu\lambda}
b_{\mu}\partial_{\nu}b_{\lambda}
\\frac{2}{g}
\left(c^2f_{12}^2-f_{20}^2-f_{01}^2
\right)-e
J_{\mu}\delta A_{\mu}
\end{equation}
where
$f_{\mu\nu}=\partial_{\mu}b_{\nu}-\partial_{\nu}b_{\mu}$
and
$g=16\pi^2\bar{\rho}/m$.

We consider a finite sample with both upper and lower
edges. If two edges are well separated, we can assume
that they are independent, and are described respectively
by that of a semi-infinite plane $y<0$ ($y>0$) for the upper
(lower) edge.
We first turn off the additional electromagnetic field $\delta A$.
The edge mode is a self-induced eigenmode of the FQH
liquid which localizes near the boundary $y=0$ as
$J_{\mu}(t,x,y)=\lambda e^{\lambda y}J_{\mu}^{(1D)}(t,x)$
for $\mu=1,2$, where $J_{\mu}^{(1D)}$ is the one-dimensional
(1D) current, and
$\lambda>0$ ($\lambda<0$) for the
upper (lower) edge.
In the absence of $\delta A$ we can assume
$J_2(t,x,y)=0$ which is
consistent with the boundary condition that the
current does not flow outside the sample. Note that only with
that condition the action remains gauge invariant
~\cite{Nagaosa95}.
Hence the current conservation reduces to
$\partial_0J_0+\partial_1J_1=0$ which is satisfied by
introducing the field $\phi^{(u,l)}$ for the upper
(lower) edge such that
$J_0^{(1D)}(t,x)=\partial_1\phi^{(u,l)}(t,x)/2\pi$ and
$J_1^{(1D)}(t,x)=\mp\partial_0\phi^{(u,l)}(t,x)/2\pi$.
We substitute these current densities into the equation
of motion for $b_{\mu}$ derived from (A4).
A solution is obtained when
$\lambda=\pm g/8\pi\nu c$ with $J_1=\mp cJ_0$, and
$(\partial_0\pm c\partial_1)J_0=0$.
The electron operator on the edges are
written in terms of $\phi^{(u,l)}$ as
\begin{equation}
\Psi^{(u,l)}(t,x,y=0)=
\sqrt{\bar{\rho}}\exp\left(\pm i\phi^{(u,l)}(t,x)/\nu\right)
\end{equation}
which guarantees the relation between the
2D current and the phase of the electron operator:
$\vec{J}^{(u,l)}(t,x,y=0)=
(\bar{\rho}/m)\nabla\phi^{(u,l)}(t,x)$
~\cite{Nagaosa94}.
Putting the 2D current into (A4),
we obtain a Lagrangian density for a 1D boson:
\begin{equation}
{\cal L}^{(1D)(u,l)} = \int dy{\cal L}^{(2D)}
= \frac{c}{8\pi\nu}
\left\{
\left({1\over c}\frac{\partial\phi^{(u,l)}}{\partial t}\right)^2+
\left(\frac{\partial\phi^{(u,l)}}{\partial x}\right)^2
\right\}
\mp\phi^{(u,l)}\bar{E_1}
\end{equation}
where
$\bar{A}_{\mu}(t,x)=\lambda\int dy
e^{\lambda y}A_{\mu}(t,x,y)$.
It should be noted that the existence of the Maxwell terms
in the Lagrangian density (A4) is essential for our
derivation.
If two edges are well separated, we can write the total 1D
Lagrangian density as a sum two chiral ones.
It will be convenient to write the 1D Euclidean Lagrangian
density as
\begin{equation}
{\cal L}_E^{(total)}=
\frac{c}{8\pi\nu}
\left\{
\left(\frac{\partial\phi^{(+)}}{\partial x}\right)^2+
\left(\frac{\partial\phi^{(-)}}{\partial x}\right)^2
\right\}
+\frac{i}{4\pi\nu}
\frac{\partial\phi^{(+)}}{\partial \tau}
\frac{\partial\phi^{(-)}}{\partial x}
\end{equation}
where
$\phi^{(\pm)}=\phi^{(u)}\pm\phi^{(l)}$.
Using the canonical quantization procedure, we can show
$[J_0^{(u,l)}(k),J_0^{(u,l)}(-k)]=\pm\nu k/2\pi$
which is consistent with the anomaly equation
$\partial_0J_0^{(1D)}+\partial_1J_1^{(1D)}
=(\nu/2\pi)E_1(t,x,y=0)$
~\cite{Nagaosa95}.

%\vfil\eject
\begin{center}
\begin{tabular}{|c|c|c|}\hline
Exponents    & (i) $\nu_1\nu_2>0$ & (ii) $\nu_1\nu_2<0$
\\ \hline
$\alpha_{11}$ & $|\nu_1|-1$          & $\nu_1 p-1$
\\ \hline
$\alpha_{22}$ & $|\nu_2|-1$          & $-\nu_2 p-1$
\\ \hline
$\alpha_{12}$ &
${|\nu_1|+\rho^2|\nu_2|\over2}-1$     &
${\nu_1-\rho^2\nu_2\over2}p-\rho q-1$
\\ \hline
Exponents    & Universal          & Non-universal
\\ \hline
\end{tabular}
\end{center}

\noindent
Table 1:
$\alpha_{IJ}$'s are listed both for
$\nu_1\nu_2>0$ (i) and for $\nu_1\nu_2<0$ (ii).
They are universal only in case (i),
otherwise depend on the interactions.
Here $p$ and $q$ are given by
$p=(v_1-v_2)/g, q=2\nu_1\nu_2g_{12}/g$ with
$g=\sqrt{(v_1-v_2)^2+4\xi}$ and $\rho=\nu_1/\nu_2$.
Later we will define $\alpha_{12}$ as $\nu_3/2-1$.

%\vspace{3cm}
\begin{center}
\begin{tabular}{|c|c|c|c|cc|}\hline
                       & \multicolumn{3}{c|}{(i) $\nu_1\nu_2>0$} &
\multicolumn{2}{c|}{(ii) $\nu_1\nu_2<0$}
\\ \hline
($\nu_1,\nu_2$)        & (${1\over5},{1\over5}$) &
(${1\over3},{1\over15}$) & (${1\over3},{1\over3}$) &
($1,-{1\over3}$)         & [with $g_{12}\rightarrow 0]$
\\ \hline
$\nu$                    & \multicolumn{2}{c|}{$2/5$} &
\multicolumn{3}{c|}{$2/3$}
\\ \hline
$\rho$                 & 1 & 5 & 1 & $-3$ & $[-3]$
\\ \hline\
$\alpha_{11}$          & $-4/5$ & $-2/3$   & $-2/3$ & $p-1>0$     & [0]
\\ \hline
$\alpha_{22}$          & $-4/5$ & $-14/15$ & $-2/3$ & $p/3-1$     & [$-$2/3]
\\ \hline
$\alpha_{12}$          & $-4/5$ & 0        & $-2/3$ & $2p-3q-1>0$ & [1]
\\ \hline
\end{tabular}
\end{center}

\noindent
Table 2:
Explicit values of $\alpha_{IJ}$'s are shown for
the specific edge constructions with filling factors
$\nu=2/5$ and $\nu=2/3$.
Note that most of the potentials are relevant.
Some exceptions are inter-channel tunnelings in the case of
$(\nu_1,\nu_2)=(1/3,1/15)$ and $(1,-1/3)$.
In both cases $\rho$ takes a large absolute value, which blocks
inter-channel tunnelings.
The other exception is the $U_{11}$ in the case of
$(\nu_1,\nu_2)=(1,-1/3)$.
The process is marginal in the absence of inter-channel
interactions, but becomes irrelevant in the presence of
infinitesimal $g_{12}$ since $p>1$.

%\vfil\eject
\begin{center}
\begin{tabular}{|c|c|c|c|c|c|c|c|c|c|}               \hline
$I\backslash j$ 
  &\ 1\ \ &\ 2\ \ &\ 3\ \ &\ 4\ \  &\ 5\ \ &
\ 6\ \ &\ 7\ \ &\ 8\ \ &\ 9\ \ \\ \hline
1 & 1 & 1 & 0 & 0 & 1 & $-1$ & 2 & 0 & 0 \\ \hline
2 & 1 & $-1$ & 1 & 1 & 0 & 0 & 0 & 2 & 0 \\ \hline
3 & 0 & 0 & 1 & $-1$ & 1 & 1 & 0 & 0 & 2 \\ \hline
\end{tabular}
\end{center}

\noindent
Table 3:
$C_{Ij}$'s are listed.
$I$ specifies the "coordinate" in the
($\theta_1, \theta_2, \theta_3$)-space, and
$j$ denotes the species of
an instanton or an anti-instanton.

%\vspace{3cm}
\begin{center}
\begin{tabular}{|c|c|c|}\hline
                  & (i) $\nu_1\nu_2>0$ &
(ii) $\nu_1\nu_2<0$
\\ \hline
$\nu_3$           & $|\nu_1|+\rho^2|\nu_2|$ &
$(\nu_1-\rho^2\nu_2)p-2\rho q$
\\ \hline
$\beta_{1,2}$      & ${1\over|\nu_1|}+{1\over|\nu_2|}-1$ &
$\left(
{1\over\nu_1}-{1\over\nu_2}
\right)p
\pm{2q\over\nu_1\nu_2}-1$
\\ \hline
$\beta_3=\beta_4$ & ${1\over|\nu_2|}+{1\over\nu_3}-1$ &
$-{p\over\nu_2}+{1\over\nu_3}-1$
\\ \hline
$\beta_5=\beta_6$ & ${1\over\nu_3}+{1\over|\nu_1|}-1$ &
${1\over\nu_3}+{p\over\nu_1}-1$
\\ \hline
$\beta_{7}$       & ${4\over|\nu_{1}|}-1$ &
${4p\over\nu_{1}}-1$
\\ \hline
$\beta_{8}$       & ${4\over|\nu_{2}|}-1$ &
$-{4p\over\nu_{2}}-1$
\\ \hline
$\beta_{9}$       & ${4\over\nu_3}-1$ &
${4\over\nu_{3}}-1$
\\ \hline
Exponents        & Universal &
Non-universal
\\ \hline
\end{tabular}
\end{center}

\noindent
Table 4:
$\beta_j$'s are listed both for
$\nu_1\nu_2>0$ (i) and for $\nu_1\nu_2<0$ (ii).
They are universal only in case (i),
otherwise depend on the interactions.

%\vfil\eject
\begin{center}
\begin{tabular}{|c|c|c|c|cc|}\hline
                       & \multicolumn{3}{c|}{(1) $\nu_1\nu_2>0$} &
\multicolumn{2}{c|}{(2) $\nu_1\nu_2<0$}
\\ \hline
($\nu_1,\nu_2$)        & $({1\over5},{1\over5})$ &
(${1\over3},{1\over15}$) & $({1\over3},{1\over3})$ &
($1,-{1\over3}$)         & [with $g_{12}\rightarrow 0]$
\\ \hline
$\nu$                    & \multicolumn{2}{c|}{$2/5$} &
\multicolumn{3}{c|}{$2/3$}
\\ \hline
$\rho$               & 1   & 5 & 1  & $-3$      & [-3]
\\ \hline
$\nu_3$              & 2/5 & 2 &2/3 & $4p-6q$   & [4]
\\ \hline
$\beta_{1,2}$        & 9 &17 & 5 &
$4p\mp6q-1$          & [3]
\\ \hline
$\beta_3=\beta_4$    & ${13\over2}^{\dagger}$ &
${29\over2}$           & ${7\over2}^{\dagger}$ &
$3p+{1\over4p-6q}-1$ & [${9\over4}$]
\\ \hline
$\beta_5=\beta_6$    & ${13\over2}^{\dagger}$ &
${5\over2}^{\dagger}$  & ${7\over2}^{\dagger}$ &
${1\over4p-6q}+p-1$  & [${1\over4}$]
\\ \hline
$\beta_{7}$          & 19 & 11 & 11 &
$4p-1$                 & [3]
\\ \hline
$\beta_{8}$          & 19 & 59 & 11 &
$12p-1$              & [11]
\\ \hline
$\beta_{9}$          & 9 & 1 & 5 &
${4\over4p-6q}-1<0$    & [0]
\\ \hline
\end{tabular}
\end{center}

\noindent
Table 5:
$\beta_j$'s are evaluated for the specific edge constructions.
Minima of $\beta_j$'s are marked by a superscript $\dagger$.
For the (1, $-$1/3) state $\beta_9<0$, i.e.
$y_9$ scales to infinity
when $g_{12}$ is finite.
The instanton picture
gives the correct temperature dependence of the tunneling conductance
only when all $y_j$'s are irrelevant,
i.e. all the original scattering potentials scale to infinity. 

%\vspace{3cm}
\begin{center}
\begin{tabular}{|l|c|c|c|}\hline
Inter-channel & \multicolumn{3}{c|}{$(\nu_1,\nu_2)$}
\\ \cline{2-4}
tunnelings    & (${1\over5},{1\over5}$) &
(${1\over3},{1\over15}$) & (${1\over3},{1\over3}$)
\\ \hline
absent        & 8 & 4 & 4 
\\ \hline
present       & 13 & 5 & 7
\\ \hline
\end{tabular}
\end{center}

\noindent
Table 6: Explicit values of $\delta_{ET}$'s
are shown both in the presence
and in the absence of inter-channel tunnelings.

%\vfil
\eject
\noindent
Fig.1

\noindent
The whole $(u_1,u_2)$-plane is divided into two domains separated
by a boundary (the dotted line), each of which has the attracive
fixed point $(0,\infty)$  (a) and $(\infty,\infty)$ (b),
respectively.
The former (a) corresponds to the picture where only quasihole
channel is reflected, and the latter (b)
to the one where both channels are
reflected.

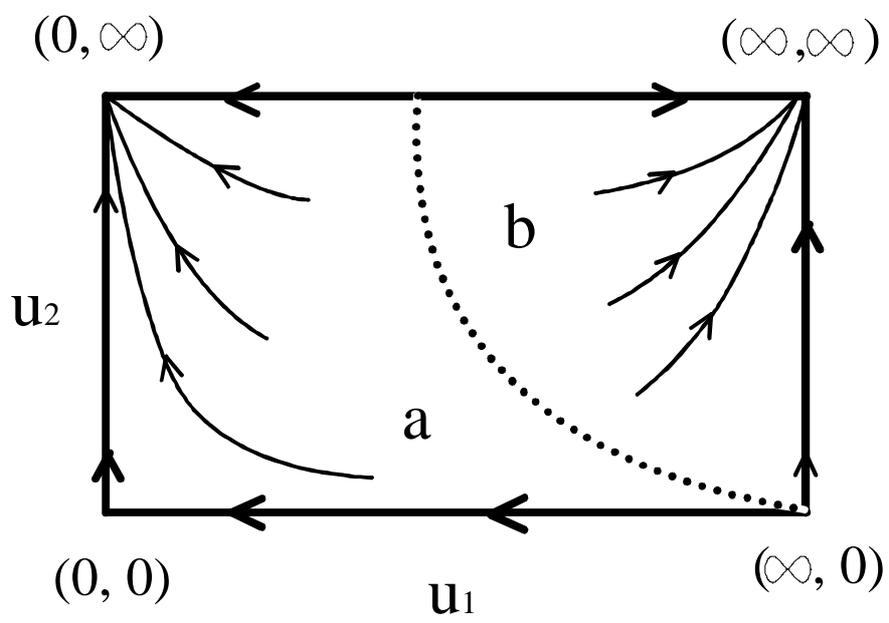
\end{document}